\documentclass[12pt,preprint]{aastex}

\def\lessim{\mathrel{\hbox{\rlap{\hbox{\lower4pt\hbox{$\sim$}}}\hbox{$<$}}}}
\def\grtsim{\mathrel{\hbox{\rlap{\hbox{\lower4pt\hbox{$\sim$}}}\hbox{$>$}}}}

\shorttitle{Photometry of VS0329+1250}
\shortauthors{Shafter et~al.}

\begin{document}


\title{Photometry of VS0329+1250:\\ A New, Short-Period SU~Ursae Majoris Star}


\author{A. W. Shafter, E. A. Coelho \& J. K. Reed}
\affil{Department of Astronomy and Mount Laguna Observatory\\
     San Diego State University\\
    San Diego, CA 92182}
\email{aws@nova.sdsu.edu,ecoelho@sciences.sdsu.edu,jreed@sciences.sdsu.edu}




\begin{abstract}

Time-resolved CCD photometry is presented of the recently-discovered
($V\simeq15$ at maximum light) eruptive variable star in Taurus,
which we dub VS0329+1250.
A total of $\sim$20~hr of data obtained over six
nights reveals superhumps in the light curves,
confirming the star as a member of the SU~UMa class of dwarf novae.
The superhumps recur with a mean period of 0.053394(7) days
(76.89~min), which represents the shortest superhump period
known in a classical SU~UMa star.
A quadratic
fit to the timings of superhump maxima reveals that the superhump
period was increasing
at a rate given by $\dot P_\mathrm{sh}\simeq(2.1\pm0.8)\times10^{-5}$ over
the course of our observations.
An empirical relation between orbital period and
the absolute visual magnitude of dwarf novae at maximum light,
suggests that VS0329+1250 lies at a distance of $\sim1.2\pm0.2$~kpc.

\end{abstract}



\keywords{binaries: novae, cataclysmic variables -
  stars: dwarf novae - stars: individual (\objectname{VS0329+1250})}


\section{Introduction}

The SU~UMa stars are a subclass of the dwarf novae that
are observed at orbital periods, $P\lessim3$~hr (for reviews
see Warner 1995, Osaki 1996). These systems
usually display two types of outburst: normal
dwarf nova outbursts, and ``superoutbursts", which are the defining
characteristic of the SU~UMa stars. Superoutbursts are usually
brighter (by $\sim0.5$~mag), and always last longer (weeks vs. days),
compared to normal outbursts. During superoutbursts, the
SU~UMa stars always display $\sim0.2$~mag modulations in their light
curves called ``superhumps", which have
been attributed to the prograde precession of an eccentric accretion
disk in these systems (Whitehurst 1988). The shortest period
SU~UMa systems that display
particularly large superoutburst amplitudes ($\grtsim6$~mag),
long intervals between eruptions ($\grtsim5$~yrs), and
light curve modulations near the orbital period early in the
outburst (so called ``early superhumps") 
are sometimes referred to as the WZ~Sge stars (Bailey 1979, Kato et~al. 2001).
 
VS0329+1250 was discovered as an optical transient
on 26 October 2006 as part of a comet and asteroid search program
(PIKA) conducted with the Crni Vrh 0.6-m Cichocki
telescope (Skvarc 2006). Follow-up observations reported in
Skvarc (2006) revealed a $\sim$0.2~mag modulation with a period of
$\sim0.0541$ days, which led to the
suggestion that the variations were superhumps, and that
the object was likely a previously unknown SU~UMa dwarf nova.

In this paper we present time-resolved V-band photometry
of VS0329+1250 originally reported in Shafter et~al. (2006)
that has enabled us to confirm the object
as an SU~UMa dwarf nova,
to estimate its distance,
and to establish a revised superhump period, which is
the shortest currently known among the classical SU~UMa stars.

\section{Observations}

A finding chart for VS0329+1250 (in outburst) is shown in Figure~1,
with the coordinates taken from Skvarc (2006).
Observations
were carried out during six nights in 2006 October and November using
the Mount Laguna Observatory 1~m reflector.
On each night a series of
exposures (typically 60~s) were taken through a Johnson-Cousins
$V$ filter (see Bessel 1990), and imaged on a Loral $2048^2$ CCD.
To increase the time-sampling efficiency, only an $800\times800$
subsection of the
full array was read out. The subsection was chosen to include
VS0329+1250 and
several relatively bright nearby stars to be used as comparison objects for
differential photometry.
A summary of observations is presented in Table~1.

The data were processed in a standard fashion (bias subtraction and
flat-fielding) using IRAF.\footnote{
IRAF (Image Reduction and Analysis Facility) is distributed by the
National Optical Astronomy Observatories, which are operated by AURA, Inc.,
under cooperative agreement with the National Science Foundation.}
The individual images were subsequently
aligned to a common coordinate system and
instrumental magnitudes for VS0329+1250
and two nearby comparison stars were then determined using the
{\it IRAF\/} APPHOT package.
Variations in atmospheric transparency were removed to first order by dividing
the flux of VS0329+1250 by a nearby comparison star located approximately
$40''$E and $20''$N of the variable (star ``C" in Fig. 1).
The $V$-band differential light curves were then placed on an
absolute scale by calibration of the comparison star against the
standard stars in Landolt (1992). The magnitude of the comparison star
is characterized by $B=15.0\pm0.1$ and $V=14.2\pm0.1$.
The calibrated light curves of VS0329+1250 are
displayed in Figure~2. In addition to the $V$-band data,
a few measurements
were obtained through a $B$ filter at the end of our $V$ sequence on
29-Oct-06 UT in order to estimate the $B-V$ color of VS0329+1250.
We find $B-V\simeq0.1\pm0.1$ at the time of superhump minimum.

\section{The Superhump Period}

During the six nights of observation, we observed a total of 16
superhump maxima. The timings of maximum light were obtained from
the point of intersection of straight line segments fitted by eye to the
ascending and descending portions of the hump profile. The accuracy
of our timings degraded over time as the hump morphology changed
from a symmetrical ``sawtooth" shape early on to a more asymmetrical and
less defined profile,
as the hump amplitude diminished.
We estimate the timings of maximum, given in Table~2, to be generally
accurate to better than $\pm2$~min, with the timings for the last night
of observation somewhat more uncertain.

A linear least-squares fit of the superhump maxima from Table~2
yields the following ephemeris for superhump maximum in VS0329+1250:
\begin{equation}
T_{\mathrm{max}} = {\rm HJD}~2,454,036.8135(4)+0.053394(7)~E.
\end{equation}
Residuals of the individual timings with respect to eqn~(1)
are included in Table~2, and are plotted as a function of cycle number
in Figure~3. Despite the considerable scatter,
a trend with cycle number is evident, which suggests that the superhump period
increased during the course of our observations.
A quadratic fit to the residuals, shown by the solid line
in Figure~3, gives
$\dot P_\mathrm{sh}\simeq(2.1\pm0.8)\times10^{-5}$.
The mean superhump period of 0.053394~d (76.89~min) is slightly
shorter than that of TSS~J022216.4+412259.9 (Imada et~al. 2006),
which had previously held the record as
the SU~UMa system with the shortest known superhump
period\footnote{The unusual dwarf novae
V485~Cen (Olech 1997)
and EI~Psc (Oemura et~al. 2002a) have shorter periods, but it is not yet clear
how they are related to the classical SU~UMa stars (e.g. Oemura 2002b).}

For many years it was thought that
the superhumps in SU~UMa stars were characterized by $\dot P_\mathrm{sh}<0$,
with the superhump period drifting lower as the
outburst evolved (Warner 1985, Patterson et~al. 1993).
However, the situation changed
about a decade ago with the discovery of an apparently
stable superhump period during an outburst of
AL~Com (Howell et~al. 1996, Patterson et~al. 1996),
and a period
increase during an outburst of SW~UMa (Semeniuk et~al. 1997;
Nogami et~al. 1998).
With the discovery of several new SU~UMa stars
in recent years, it has now been clearly established that
among the shortest period systems, such as VS0329+1250,
increases in the superhump period are usually observed
(e.g. Imada et~al. 2005, Uemura et~al. 2005, and references therein).
It has been proposed that
the period increases seen in the shortest period
SU~UMa stars with extreme mass ratios
may result from a tidal resonance radius well inside the primary's
Roche lobe that enables the eccentric disk wave to propagate outwards
(Baba et~al. 2000).

The superhump periods in SU~UMa stars are usually longer
than the orbital periods,
with the ``period excess", defined as
$\epsilon\equiv(P_\mathrm{sh}-P_\mathrm{orb})/P_\mathrm{orb}$, generally
increasing linearly with $P_\mathrm{orb}$ (Stolz \& Schoembs 1984, Patterson
1998). The relationship becomes less well defined at the shortest orbital
periods, where $\epsilon$ typically lies in the range
$0.01\lessim~\epsilon~\lessim0.03$.
At present, the period excess for VS0329+1250
cannot be computed precisely because
the orbital period of the system has not yet been directly measured.
Nevertheless, if VS0329+1250
has a period excess typical of SU~UMa stars with similar superhump
periods, we estimate that the orbital period, $P_\mathrm{orb}\simeq76$~min.
Once the orbital period is measured, and a precise value of $\epsilon$
determined, it will become possible to begin to explore the structure
and evolutionary state of the VS0329+1250 system using existing
empirical and theoretical relationships between
the period excess and mass ratio (e.g. Patterson 2001,
Patterson et~al. 2005, Pearson 2006).

\section{The Distance to VS0329+1250}

We can make a rough estimate of the distance to VS0329+1250 by appealing
to the empirical relation between the orbital period and absolute
magnitude of dwarf novae at maximum light
given by Warner (1987). For normal dwarf
nova eruptions Warner finds $M_V(\mathrm{max})=5.64-0.259~P(\mathrm{hr})$.
Given a period
of 1.27~hr, we estimate $M_V(\mathrm{max})\simeq5.3\pm0.3$,
where the assumed uncertainty reflects the effect of the
unknown orbital inclination on the absolute magnitude of the system.
Since the outburst
of VS0329+1250 presented here is the only recorded outburst of the system,
the peak magnitude of a normal outburst (assuming VS0329+1250 exhibits
normal outbursts) is not known. For our purposes, we will assume
a normal outburst magnitude that
is 0.5~mag fainter than the
superoutburst maximum, which is typical of SU~UMa stars.
Thus, we adopt $m_V(\mathrm{max})=15.7$ for VS0329+1250 based upon
the observed superoutburst maximum of $m_V=15.2$ (Skvarc 2006).
The high Galactic latitude of VS0329+1250 ($|b|\simeq35^{\circ}$) suggests that
the interstellar extinction to the object is likely to be negligible.
Thus, we estimate that VS0329+1250 lies at a distance,
$d\simeq1.2\pm0.2$~kpc.
The fact that VS0329+1250 is not
visible on the POSS or the DSS reveals that the eruption
amplitude, $A\grtsim6$~mag, which is typical of the shortest period
SU~UMa stars (WZ~Sge stars), and suggests that $M_V(\mathrm{min})\grtsim11$.

\section{Conclusions}

We have presented six nights of time-resolved V-band CCD photometry of
the eruptive variable star, VS0329+1250,
recently discovered by Skvarc (2006). Our principal conclusions are
as follows:

1) Superhumps are seen in
the light curves of VS0329+1250 that recur with a period
$P_\mathrm{sh}=0.053394(7)$~d, clearly establishing the system
as a member of the SU~UMa class of dwarf novae.
The superhump period is the shortest
of any known SU~UMa system, with the possible exception of
the dwarf novae V385 Cen and EI Psc,
whose relationship to the classical SU~UMa systems is unclear.

2) VS0329+1250 faded by about 0.08 mag per day from a mean magnitude of
$V\simeq15.6$ on our first night of observation (2 days post discovery)
to $V\simeq16.1$ six
nights later, at which time the superhumps had nearly disappeared.

3) The superhump timings are consistent with an increasing period over the
week spanned by our observations,
with $\dot P_\mathrm{sh}\simeq(2.1\pm0.8)\times10^{-5}$.

4) Using Warner's (1987) empirical relation between the absolute visual
magnitude of dwarf novae at maximum light and orbital period,
we estimate a distance to VS0329+1250 of $\sim1.2\pm0.2$~kpc.

Future observations will be required to firmly establish the
nature of VS0329+1250, and to determine if it
should be included as a member of the WZ~Sge stars.
Specifically, the system
should be monitored for subsequent outbursts
that will establish the outburst recurrence time.
More extensive photometry during future superoutbursts should
help to better characterize the stability of the superhump period.
Finally, a single deep image of the field should be attempted
to measure the quiescent magnitude
of VS0329+1250, and thereby establish its outburst amplitude.



\acknowledgments
We thank the referee for insightful comments on the original manuscript.
This research was partially supported by NSF grant AST-0607682.

\clearpage

\begin{deluxetable}{ccccc}
\tablenum{1}
\tablewidth{0pt} 
\tablecolumns{5}
\tablecaption{Summary of Observations}
\tablehead{\colhead{} 					&	 
           \colhead{UT Time} 				& 
	   \colhead{Time Resolution\tablenotemark{a}} 	&
	   \colhead{Number of} 				&
	   \colhead{} 					\\ 
	   \colhead{UT Date} 				& 
	   \colhead{(start of observations)} 		&
	   \colhead{(sec)} 				& 
	   \colhead{Exposures} 				& 
	   \colhead{Filter}				}
\startdata
2006 Oct 28 &06:30:02.5 &38.32 &100 &V\\
2006 Oct 29 &05:33:00.0 &65.43 &372 &V\\
2006 Oct 29 &12:20:00.0 &65.43 &3 &  B\\
2006 Oct 30 &04:45:08.4 &65.43 &330 &V\\
2006 Nov 01 &05:12:00.0 &65.43 &92  &V\\
2006 Nov 02 &07:45:00.0 &65.43 &130 &V\\
2006 Nov 03 &06:21:00.0 &65.43 &130 &V\\
\enddata
\tablenotetext{a}{Mean time interval between exposures (integration time plus 
readout time)}
\end{deluxetable}

\clearpage

\begin{deluxetable}{lcr}
\tablenum{2}
\tablecaption{Superhump Timings}
\tablewidth{0pt}
\tablehead{	\colhead{HJD (peak)} 	& 
		\colhead{Cycle Number}		& 
		\colhead{$O-C$} 		\\
		\colhead{(2,450,000+)} 		& 
		\colhead{$(E)$} 		& 
		\colhead{($\times10^{-3}$~day)}	} 
\startdata
4036.8146\dots &  0  &  $1.075$  \\
4037.7749\dots &  18 &  $0.297$  \\
4037.8272\dots &  19 &  $-0.847$  \\
4037.8809\dots &  20 &  $-0.471$  \\
4037.9355\dots &  21 &  $0.664$  \\
4037.9885\dots &  22 &  $0.280$  \\
4038.7357\dots &  36 &  $-0.070$  \\
4038.7894\dots &  37 &  $0.295$  \\
4038.8432\dots &  38 &  $0.681$  \\
4038.8947\dots &  39 &  $-1.223$  \\
4038.9492\dots &  40 &  $-0.087$  \\
4040.7634\dots &  74 &  $-1.265$  \\
4041.8311\dots &  94 &  $-1.501$  \\
4041.8856\dots &  95 &  $-0.366$  \\
4042.7940\dots & 112 &  $0.291$  \\
4042.8493\dots & 113 &  $2.247$  \\
\enddata
\label{time}
\end{deluxetable}

\clearpage

\begin{figure}
\epsscale{0.80}
\plotone{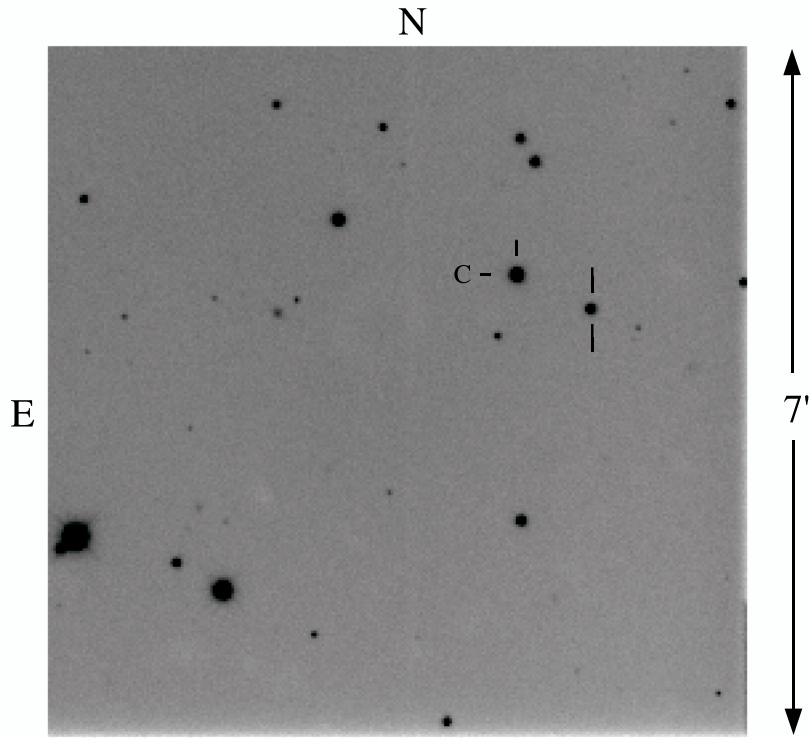}
\caption{The finding chart for VS0329+1250. The coordinates of
          VS0329+1250 are RA. = 03:29:12.26, DECL. = +12:50:17.6
          (equinox 2000.0) as given in Skvarc (2006).
          The comparison star used to calibrate our data is located
          $\sim40''$E and $\sim20''$N of VS0329+1250, and is marked
          as star ``C".}
\end{figure}

\clearpage

\begin{figure}
\epsscale{0.80}
\plotone{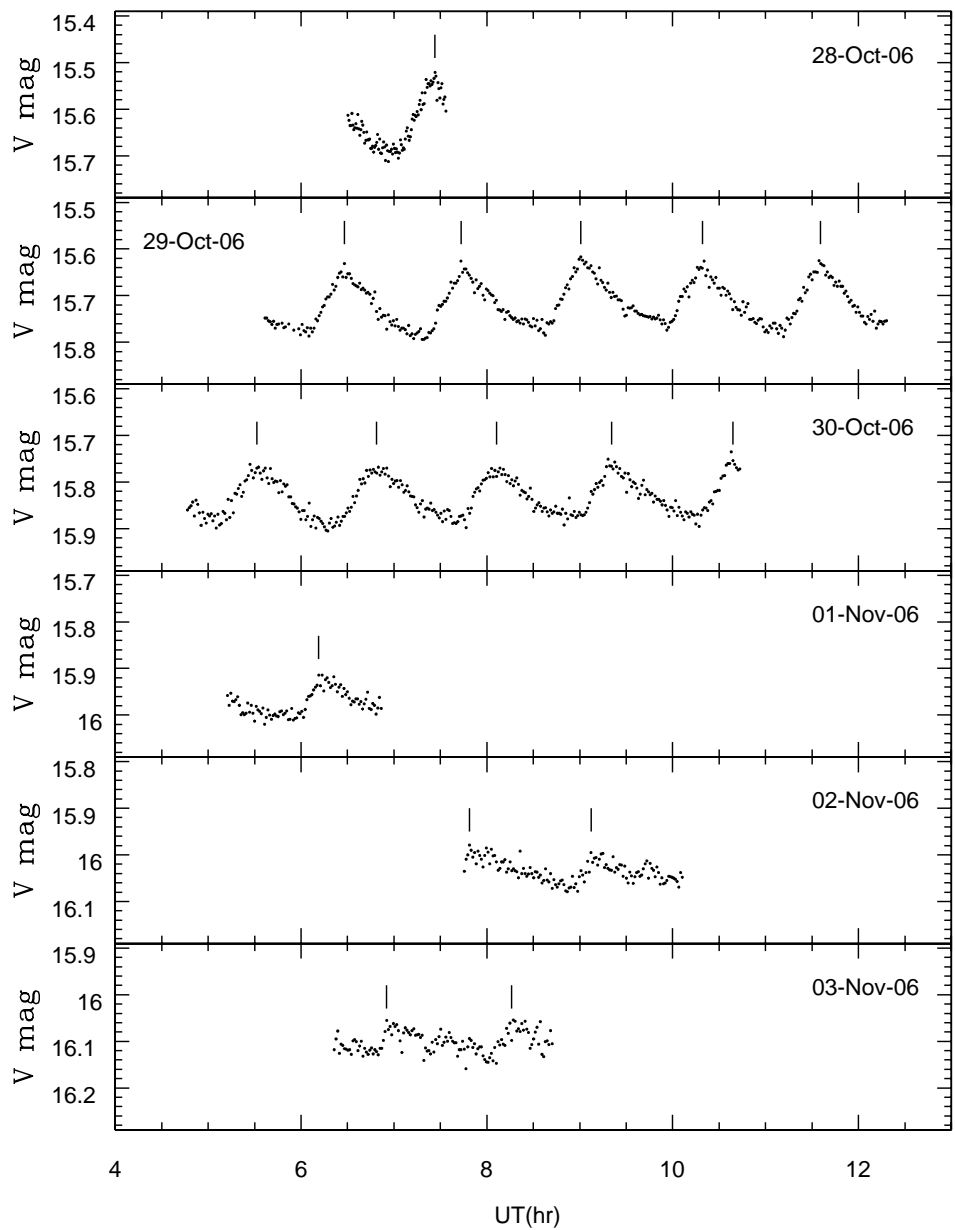}
\caption{The $V$-band light curves of VS0329+1250.
          Superhumps are clearly evident, with the hump amplitude
          gradually decreasing and changing shape
          as the system fades over the seven days
          spanned by our data.
          The measured times of superhump maximum are indicated by
          the vertical tick marks. The first timing on 02-Nov-06, and
          both timings on 03-Nov-06 are relatively uncertain due to
          poor coverage of hump maximum, and decreasing signal-to-noise.}
\end{figure}

\clearpage

\begin{figure}
\epsscale{0.80}
\plotone{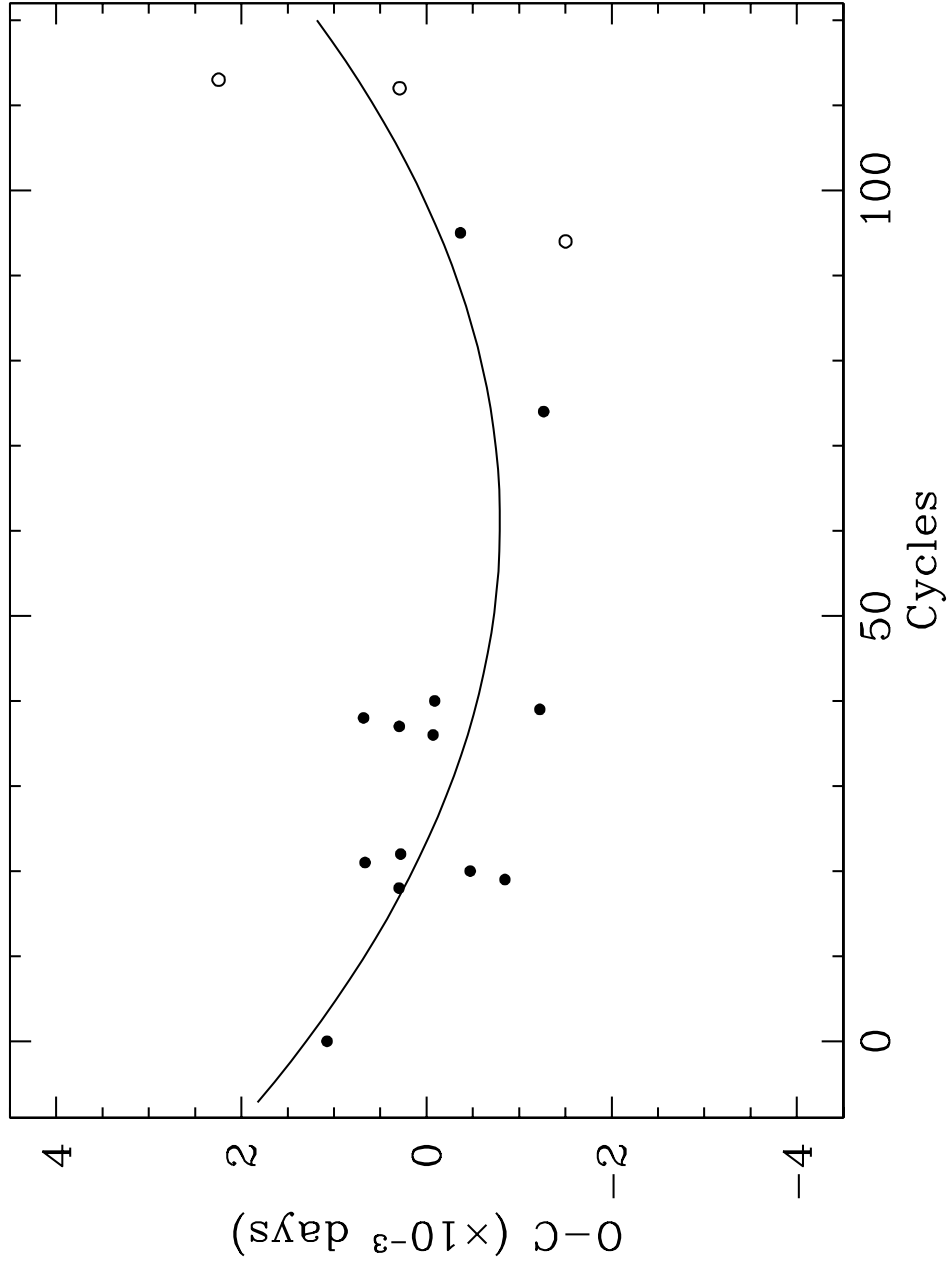}
\caption{The residuals of the observed times of superhump maximum with respect
         to the times predicted from eqn~(1) are plotted as a function of
         superhump cycle number. The open circles represent the superhump
         timings with relatively large uncertainties.
         These data suggest an increase
         in the superhump period, with
         the fit shown by the solid line corresponding to
         $\dot P_\mathrm{sh}\simeq(2.1\pm0.8)\times10^{-5}$.}
\end{figure}

\end{document}